\documentclass[final,3p,times,twocolumn,preprint]{elsarticle}
\usepackage{amssymb}
\usepackage{amsmath}
\usepackage{verbatim}
\usepackage{amsfonts}
\usepackage{hyperref}

\allowdisplaybreaks

\biboptions{sort&compress,merge}

\begin{document}

\begin{frontmatter}

\title{Branching ratio measurements and isospin violation in $B$-meson decays}
\author[TUM]{Martin Jung}
\ead{martin.jung@tum.de}
\address[TUM]{TU Munich IAS and Universe Cluster of Excellence, Boltzmannstr.2, D-85748 Garching, Germany}

\begin{abstract}
The approximate symmetry of the strong interactions under isospin transformations is among the most precise tools available to control hadronic matrix
elements. It is crucial in extracting fundamental parameters, but also provides avenues for the search of phenomena beyond the Standard Model. The
precision of the resulting predictions requires special care when determining the quantities they are to be tested with. Specifically, in the
extraction of branching ratios often isospin symmetry is assumed at one point or another implicitly, implying a significant bias for precision
analyses. We extract a bias-free value for the production asymmetry between charged and neutral $B$ meson pairs at $B$ factories and discuss its
consequences for the determination of branching fractions generally, and isospin-violating observables like the rate asymmetries in $B\to J/\psi K$ or
$B\to K^*\gamma$ decays specifically.
\end{abstract}

\end{frontmatter}

\section{Introduction}
Isospin symmetry is a widely used approximation in particle physics. This is typically justified, given that apart from the weak interactions it is
broken only by the difference of the up- and down-quark masses and charges, yielding generally corrections at the percent level. This enables
determinations of fundamental parameters as well as searches for phenomena beyond the Standard Model (SM), commonly called \emph{new physics} (NP). 

While the uncertainty related to isospin breaking can often be neglected compared to that from other sources, in precision measurements care must be
taken to account for it properly; this is complicated by the fact that the assumption of isospin symmetry often enters implicitly in input quantities.
A prime example is the production asymmetry of charged and neutral $B$ meson pairs at $B$ factories. It is commonly either assumed to vanish or
determined using measurements assuming in turn isospin symmetry for the weak decay in question. While a priori the latter approach seems reasonable
given that isospin breaking in the $\Upsilon(4S)$ decay from electromagnetic interactions is parametrically enhanced by the
small velocity $v$ of the $B$ mesons as $\pi^2/v\sim\mathcal O(100)$~\cite{Atwood:1989em}, available
data~\cite{Alexander:2000tb,*Athar:2002mr,Hastings:2002ff,Aubert:2004ur,Aubert:2005bq} indicate that the breaking is actually comparable to that in
weak decays. In any case, input values extracted assuming isospin symmetry in weak decays can generally not be used in experimental analyses testing
this assumption. Apart from the immediate consequences for analyses dealing explicitly with isospin breaking, the production asymmetry more generally
affects most branching ratio results from $B$ factories and also from hadron colliders where absolute branching fractions are necessary as inputs for
the normalization modes.

In the next section we show how to circumvent these problems by determining a value for the production asymmetry that is not affected by isospin
breaking. This is followed in Sec.~\ref{sec::pheno} by the study of a selection of phenomenological consequences, addressing specifically $b\to
s\bar c c$ and $b\to s\gamma$ transitions. We conclude in Sec.~\ref{sec::conclusions}.

\section{Extraction of $\boldsymbol{f_{+-}/f_{00}}$ without bias}
The relative production fraction of charged ($f_{+-}$) and neutral $B$ mesons ($f_{00}$) at $B$ factories is a crucial issue, especially when
isospin is to be tested with the results. Theoretical predictions for this quantity are difficult: while estimates for point-like $B$ mesons
predicted a large asymmetry~\cite{Atwood:1989em,Kaiser:2002bm}, model calculations~\cite{Byers:1990rd,*Lepage:1990hm,*Dubynskiy:2007xw} indicate that
the meson and vertex structures as well as strong rescattering phases could suppress the net effect. The precision of these calculations does not
(yet) match the experimental one, therefore we will concentrate on the experimental determinations in the following. 

The value most commonly used for the production asymmetry stems from the heavy flavor averaging group (HFAG), obtained as an average of various
measurements, $r_{+0}\equiv f_{+-}/f_{00}=1.058\pm0.024$~\cite{Amhis:2014hma}, about $2.4\sigma$ from unity. However, in determinations of branching
ratios typically still $r_{+0}=1$ is assumed~\cite{Agashe:2014kda}. More importantly, as also pointed out in Ref.~\cite{Amhis:2014hma}, most of the
values entering the average of $r_{+0}$ actually assume isospin symmetry for the weak decays under consideration in their analyses, \emph{e.g.}
$\Gamma(B^0\to J/\psi K^0)=\Gamma(B^+\to J/\psi K^+)$. Using the resulting value to extract information on isospin breaking in weak decays would
therefore be circular. Nevertheless, the above value indicates that the production asymmetry could be sizable, rendering its extraction without the
assumption of isospin symmetry mandatory.

One analysis by BaBar~\cite{Aubert:2005bq} uses a ratio of singly and doubly tagged $B$ decays, a method introduced in the context of resonant charm
production~\cite{Baltrusaitis:1985iw}. This technique avoids assumptions regarding isospin, yielding $f_{00}=0.487\pm0.010\pm0.008$. Assuming
$f_{00}+f_{+-}=1$, in accordance with ${\rm BR}(\Upsilon(4S)\to~{\rm non-}B\bar B)\leq 4\%~(95\%~{\rm CL})$~\cite{Barish:1995cx} and the fact that no
other decay mode has been observed with a branching fraction larger than $\sim 10^{-4}$~\cite{Amhis:2014hma}, this measurement corresponds to
$r_{+0}=1.053\pm 0.054$. Note that a significant contribution from non-$B\bar B$ events would reduce this value. Potential corrections to
the simple relation $N_d\sim BR(B\to D^*\ell\nu)^2$, \emph{e.g.} from CP violation in mixing, enter the expression at the negligible level of
$\lesssim 10^{-5}$.

Inclusive measurements are less sensitive to isospin breaking, since it is additionally suppressed by $1/m_b^2$~\cite{Gronau:2006ei};
importantly, this is even true for NP contributions. Therefore the Belle measurement using inclusive semileptonic decays~\cite{Hastings:2002ff},
\emph{i.e.} the assumption $\Gamma(B^-\to X \ell)=\Gamma(B^0\to X\ell)$, is unaffected by isospin breaking at the required level. Its uncertainty is
dominated by the lifetime ratio of neutral and charged $B$ decays; updating it to the present world average~\cite{Amhis:2014hma} yields
$r_{+0}=1.00\pm0.03\pm0.04$, where the systematic uncertainty is now a sum of several similarly large contributions.

Combining these two measurements and adding statistical and systematic uncertainties in quadrature, we obtain 
\begin{equation}\label{eq::r}
r_{+0}\equiv \frac{f_{+-}}{f_{00}} = 1.027\pm0.037\,,
\end{equation} 
compatible with unity. While less precise than the ``standard'' HFAG average given above, only this value can be used to investigate isospin breaking
without a significant bias. 

Both analyses used in the average have been performed with a small fraction of the corresponding full datasets, leaving room for significant
improvement despite potential systematic limitations: already repeating the BaBar analysis with the full dataset would reduce the total uncertainty
for $f_{00}$ to below $0.8\%$ (corresponding to $\sim 3\%$ on $r_{+0}$ when assuming $f_{+-}+f_{00}=1$, better than the current average in
Eq.~\eqref{eq::r}).\footnote{This is a conservative estimate, taking only the improved determination of the number of $B\bar B$ pairs $N_{B\bar B}$
into account and scaling the statistical uncertainty with the luminosity. Additional improvements are expected, \emph{e.g.} from the improved
knowledge of $B\to D^*\pi\ell\nu$.} The analysis could also be carried out with the available Belle dataset, and even further improved with Belle~II
data. Additional analyses not relying on isospin symmetry are therefore promising and necessary for many precision tests of the SM.

Given the very high expected luminosity at Belle~II, one could additionally consider using the modes $\bar B^{0,-}\to D^{*\,+,0}(\to
D^{+,0}\pi^0)\ell\bar\nu$:
while these modes have a smaller reconstruction efficiency, they have the advantage of allowing for the
determination of both, $f_{00}$ and $f_{+-}$. This would (i) be the first direct determination of $f_{+-}$, (ii) determine $r_{+0}$ as a double ratio
where $N_{B\bar B}$ and possibly other systematic effects cancel, and (iii) allow for an experimental test of the assumption $f_{+-}+f_{00}=1$. Given
that the latter relation constitutes the main theoretical assumption in the present determination of $r_{+0}$, it is important that also the analysis
in Ref.~\cite{Barish:1995cx} could be improved upon already with existing data.

The relations between the values for branching fractions given for $r_{+0}=1$ and the ones including the correction factor for the production
asymmetry are readily obtained as
\begin{align}
\left.{\rm \overline{BR}}(B^{+/0}\to X)\right|_{\rm corr} \hspace{-3mm}\equiv c_{+/0}\left.{\rm \overline{BR}}(B^{+/0}\to X)\right|_{r_{+0}=1}\,,
\end{align}
where the bars denote CP averages and the correction factors are $c_+=(1+1/r_{+0})/2$ and $c_0=(1+r_{+0})/2$, $c_0/c_+=r_{+0}$.

\section{Phenomenological consequences\label{sec::pheno}}

Isospin breaking in $B$ decays is typically discussed using rate asymmetries $A_I$ (also sometimes denoted $\Delta_{0-}$), 
\begin{equation}\label{eq::AI}
A_I(X)\equiv \frac{\bar\Gamma(B^0\to X_d^0)-\bar \Gamma(B^+\to X_u^+)}{\bar\Gamma(B^0\to X_d^0)+\bar \Gamma(B^+\to X_u^+)}\,. 
\end{equation}
The additional uncertainty stemming from including the production asymmetry explicitly instead of setting it naively to unity is approximately $\delta
r_{+0}/2\sim2\%$; this is therefore the present sensitivity limit which could however be improved upon by additional measurements. 
Any branching-ratio measurement at a comparable level of precision is affected by $r_{+0}$; apart from the examples given below, its effect should
also be included for instance in the extraction of $|V_{cb}|$.\footnote{A fit as used in Ref.~\cite{Amhis:2014hma} to extract $|V_{cb}|$ is beyond
the scope of this work. A first estimate does not yield a large shift compared to other uncertainties, related to the fact that the relevant branching
ratios are proportional to $|V_{cb}|^2$.} It is important to note that isospin symmetry does not predict the rate asymmetry to vanish necessarily:
in general several isospin amplitudes contribute, and while each of them will be related by isospin symmetry, different combinations can enter the two decay amplitudes in question. However, there are various examples where $A_I=0$
does hold to an approximation which is at least as good as the assumption of isospin symmetry itself; these are then dubbed \emph{quasi-isospin
relations}~\cite{Jung:2014jfa}. A well-known class of examples are processes dominated by $b\to c\bar cs$ transitions, like $B\to J/\psi K$. A sizable
rate asymmetry could in these cases indicate NP with a specific isospin structure, \emph{e.g.} $\Delta I=1$ for $B\to J/\psi K$.

\subsection{$b\to c\bar cs$ transitions}

We start by considering the branching ratios for $B\to J/\psi K$, often used as normalization modes and entering analyses of penguin pollution in
$B\to J/\psi K_S$. Using the values for the branching ratios given in Ref.~\cite{Agashe:2014kda} yields the rate asymmetry
$A_I(B\to J/\psi K)|_{r_{+0}=1}=-0.044\pm0.024$, approximately $2\sigma$ from zero. While this is not very significant, it has been cause for
speculation regarding possible NP or enhanced QCD contributions~\cite{Feldmann:2008fb,*Jung:2009pb,Jung:2012mp,Ligeti:2015yma}. 

In addition to combining the appropriate value for the production asymmetry from Eq.~\eqref{eq::r} with the world averages for the individual
branching ratios~\cite{Agashe:2014kda}, we recast another BaBar measurement for the production asymmetry using $B\to J/\psi K$~\cite{Aubert:2004ur}
into\footnote{We do not consider the correlations with the averages for the individual branching ratios. The BaBar measurements entering there have
been obtained with a larger dataset ($\sim 1.5\times$), are dominated by different uncertainties and averaged with the results from other
experiments.}
\begin{align}
\frac{f_{+-}}{f_{00}}\frac{{\rm \overline{BR}}(B^+\to J/\psi K^+)}{{\rm \overline{BR}}(B^0\to J/\psi K^0)} = 1.090\pm 0.045\,.
\end{align}
The combination of these ingredients yields
\begin{align}
{\rm \overline{BR}}(B^+\to J/\psi K^+) &= (9.95\pm0.32)\times 10^{-4}\quad{\rm and}\nonumber\\
{\rm \overline{BR}}(B^0\to J/\psi K^0) &= (9.08\pm0.31)\times 10^{-4}\,,\label{eq::BRK}
\end{align}
with an accidentally small correlation of below $1\%$. Note that the uncertainties remain basically identical compared to the values in
Ref.~\cite{Agashe:2014kda}, despite including now the uncertainty from the production asymmetry, and the central values are closer than before. This
affects the rate asymmetry, which is now given as
\begin{equation}\label{eq::AIK}
A_I(B\to J/\psi K)=-0.009\pm0.024\,,
\end{equation}
showing no sign of isospin violation in these decays. This value could be used to determine the relative production fraction $f_u/f_d$ of charged
and neutral $B$ mesons at hadron colliders.

Interestingly, the isospin asymmetry for $B\to J/\psi\pi$ tests specific contributions that are also related to the ``penguin pollution'' in $B\to
J/\psi K_S$~\cite{Jung:2012mp,Ligeti:2015yma}. The approximate $SU(3)$ relation~\cite{Jung:2012mp} 
\begin{equation}
\frac{A_I(B\to J/\psi\pi)}{A_I(B\to J/\psi K)}\approx \frac{1}{\lambda^2}\,,
\end{equation}
where $\lambda\approx0.2$ denotes the Wolfenstein parameter~\cite{Wolfenstein:1983yz}, yields a strong relative enhancement in $B\to J/\psi\pi$. The
determination of $A_I(B\to J/\psi\pi)$ is presently complicated by the fact that the two most precise measurements for $r_{\pi
K}={\rm\overline{BR}}(B^+\to J/\psi \pi^+)/{\rm \overline{BR}}(B^+\to J/\psi K^+)$~\cite{Aaij:2012jw,Aubert:2004pra} are incompatible. Using the PDG
averages for $r_{\pi K}$ (including a scale factor of 3.2) and ${\rm \overline{BR}}(B^0\to J/\psi\pi^0)$~\cite{Agashe:2014kda}, together with
Eqs.~\eqref{eq::r} and~\eqref{eq::BRK}, we obtain\footnote{Note that $2{\rm\overline{BR}}(B^0\to J/\psi\pi^0)$ is used in this case for
Eq.~\eqref{eq::AI}.}
\begin{equation}\label{eq::AIpi}
A_I(B\to J/\psi \pi)=-0.02\pm0.07\,,
\end{equation}
well compatible with zero.\footnote{Excluding one of the incompatible results yields $A_I(B\to J/\psi \pi)=0.00\pm0.05$ (excluding
\cite{Aubert:2004pra}) and $A_I(B\to J/\psi \pi)=-0.15\pm0.06$ (excluding \cite{Aaij:2012jw}); the latter value would indicate the presence of this
contribution, while still implying an isospin asymmetry of below $1\%$ in $B\to J/\psi K$.} The determination of $r_{\pi K}$ to resolve this
tension and an improved determination of ${\rm \overline{BR}}(B^0\to J/\psi\pi^0)$ at Belle~II will provide important information to restrict penguin pollution further.

Finally, a possible violation of a quasi-isospin sum rule in $B\to DD$ decays measured in Ref.~\cite{Aaij:2013fha} has recently been discussed in
Ref.~\cite{Jung:2014jfa}. Here we only point out that if this measurement were to be performed with even better precision, the relative production
fraction $f_u/f_d$ would have to be taken into account explicitly.

\subsection{$b\to s\gamma$ transitions}

One of the most precisely measured rate asymmetries is the one in $B\to K^*\gamma$, where PDG averages the BaBar~\cite{Aubert:2009ak} and
Belle~\cite{Nakao:2004th} measurements to $A_I^{\rm PDG}=0.052\pm0.026$. It is predicted in the SM to be around
$5\%$~\cite{Kagan:2001zk,*Bosch:2001gv,*Beneke:2001at,*Beneke:2004dp,*Ball:2006eu,*Lyon:2013gba}, despite (linear) $1/m_b$ suppression, due to
annihilation contributions enhanced by the ratio of Wilson coefficients $C_1/C_7$. Form factor uncertainties cancel largely, allowing for a more precise prediction than for
\emph{e.g.} the individual branching ratios. Apart from being an interesting test of QCD dynamics, this observable also yields important information
for NP, since it provides complementary information on the coefficient of the photonic penguin operator $\mathcal O_7$, as \emph{e.g.} emphasized in
Refs.~\cite{Kagan:2001zk,Xiao:2003vq,*Ahmady:2005nc,*Ahmady:2006yr,*Mahmoudi:2009zx,*DescotesGenon:2011yn,*Jung:2012vu}. Adapting the experimental
results to the present value for the lifetime ratio and using Eq.~\eqref{eq::r}, we obtain the average
\begin{align}\label{eq::AIK*}
A_I(B\to K^*\gamma) = 0.042\pm0.032\,,
\end{align}
consistent with zero as well as the prediction from QCD factorization. In this case the correction is quite small and shifts the central value in the
opposite direction, owing to the fact that the BaBar collaboration included the production asymmetry as measured at the time in their measurement, but
again making implicit assumptions on isospin breaking in other modes in the process.

For the inclusive decay $B\to X_s\gamma$, the isospin violation is again expected to be additionally suppressed. The BaBar measurement in
Ref.~\cite{Aubert:2005cua}, dominating the world average, reads with our ratios for lifetimes and production fractions
\begin{equation}\label{eq::AIXs}
A_I(B\to X_s\gamma) = -0.001(58)(5)(19)\,,
\end{equation}
with a largely dominating statistical uncertainty given first and a very small systematic one, given second. The third uncertainty is due to the
production asymmetry, which could potentially even be determined using this measurement, given that also potential NP spectator interactions receive
the additional suppression. However, due to the experimentally necessary cuts the measurement is not fully inclusive, and the other methods mentioned
above are certainly preferable given that they rely on tree-level decays.

\section{Conclusions\label{sec::conclusions}}
We analyzed in detail the determination of the production asymmetry between charged and neutral $B$ meson pairs at $B$ factories and some of its
phenomenological consequences. Contrary to early estimates, this asymmetry is comparable in size to potential isospin-violating effects in weak
decays, which requires accounting for both simultaneously in precision analyses. Here care has to be taken not to use the assumption of isospin
symmetry implicitly, in order not to spoil the resulting precision. The phenomenological results with present data neither indicate a significant
production asymmetry, nor unexpectedly large rate asymmetries, \emph{cf.} Eqs.~\eqref{eq::r},\eqref{eq::AIK},\eqref{eq::AIpi}-\eqref{eq::AIXs}.
However, for precise measurements the size of the correction can be relatively substantial, as demonstrated for $B\to J/\psi K$ decays, \emph{c.f.}
Eq.~\eqref{eq::BRK}. This shows the importance of an improved determination of $r_{+0}$ as well as proper application, especially in light of the
expected precision results from LHCb and Belle~II. In addition to improving the existing analyses, we proposed to use $\bar B^{0,-}\to D^{*\,+,0}(\to
D^{+,0}\pi^0)\ell\bar\nu$ decays, which allow to determine both production fractions directly and thereby also the amount of $\Upsilon$ decays into
non-$B\bar B$ states.

\section*{Acknowledgments}
I would like to Danny van Dyk and the Belle members in Munich for useful discussions and Tim Gershon for helpful comments on the
manuscript.
This work was performed in the context of the ERC Advanced Grant project 'FLAVOUR' (267104) and was supported in part by the DFG cluster of excellence 'Origin and Structure of the Universe'.

\bibliography{refs}

\end{document}